\documentstyle[epsfig]{europhys}

\def\drm{{\rm d}}

\newif\ifboo \boofalse

\begin{document}
%
%
\euro{48}{5}{486-490}{1999}
\Date{1 December 1999}
\shorttitle{D. A. GARANIN, NEW INTEGRAL RELAXATION TIME }
\bibliographystyle{prsty}
\title{New integral relaxation time for thermal activation of magnetic particles}
\author{
D. A. Garanin\inst{1}
\footnote{www.mpipks-dresden.mpg.de/$\sim$garanin/; 
garanin@mpipks-dresden.mpg.de
}
} 
\institute{
     \inst{1} Max-Planck-Institut f\"ur Physik  komplexer Systeme,
N\"othnitzer Str. 38, D-01187 Dresden, Germany 
}
%
%
\rec{10 May 1999}{28 September 1999}
%
\pacs{
\Pacs{05}{40.-a}{Fluctuation phenomena, random processes, noise, and Brownian motion}
\Pacs{75}{50.Tt}{Fine-particle systems}
}
\maketitle
\vspace{-1cm}
\begin{abstract}
The integral relaxation time for the {\em difference of numbers of particles in the two
potential wells} (IRT-$N$) for double-well classical spin systems is introduced.
For the uniaxial symmetry, it is given by a quadrature.
Unlike the previously introduced integral relaxation time for the magnetization,
the IRT-$N$ at low temperatures describes the rate of crossing the barrier under all conditions, including
the strongly biased case.
In the high-barrier case, the new integral relaxation time approaches the
inverse of the lowest eigenvalue of the Fokker-Planck equation $\Lambda_1$.
It can be more conveniently found by numerical methods than the latter.
\end{abstract}

\vspace{-1cm}

The rate of thermal activation $\Gamma$ of a dynamical system over a potential barrier
$\Delta U$ between two wells is exponentially small in the low-temperature or high-barrier
limit: $\Gamma =\Gamma_0 \exp(-\Delta U /T)$.
In this limit, all definitions of the rate yield the same result.
Analytical methods based on solving the Fokker-Planck equation (FPE) for the
flow of the probability current over the barrier ware developed by Kramers \cite{kra40}
for the particles (see ref.\ \cite{haetalbor90} for a review) and by Brown \cite{bro6379} for magnetic particles (classical
spins).
In situations where analytical solution is problematic, the thermal activation
rate $\Gamma$ was usually identified with the lowest eigenvalue $\Lambda_1$ of the
Sturm-Liouville equation corresponding to the FPE and $\Lambda_1$ was calculated
numerically \cite{aha6469}.
In the high-barrier case, $\Lambda_1$ is exponentially small and describes the
slow process of the exchange of particles between the two wells, whereas all other
$\Lambda_i$, $i=2,3,\ldots$ are not exponentially small and correspond to 
intrawell processes.
For spin problems in the high-temperature limit $\Delta U /T \ll 1$, the lowest eigenvalue
$\Lambda_1$ describes relaxation of the magnetization and it is, again, 
more important than other eigenvalues describing relaxation of higher
 moments of the distribution function.
At intermediate temperatures $\Delta U /T \sim 1$, the exclusive role of $\Lambda_1$
in the overbarrier relaxation is no longer obvious, since other eigenvalues are
of the same order as $\Lambda_1$ and they can make contribution to this process,
too.
In addition, $\Lambda_1$ cannot be calculated analytically in this temperature
range even for the simplest models, such as that of uniaxially anisotropic spins
with or without a longitudinal field.

In fact, at intermediate temperatures the overbarrier escape rate is not
unambiguously defined, and one can use various alternative definitions of
$\Gamma$.
One of the quantities measuring the escape rate is the mean first-passage time
(MFPT)
\cite{kle52,ris89} which for the simplest models can be calculated analytically in the
whole range of temperatures.
Another useful quantity which is more appealing to physicists is the integral
relaxation time (IRT) $\tau_{\rm int}$ which is defined by the area under the relaxation curve for the
magnetization or polarization after an infinitesimal instantaneous change of the
applied field \cite{sza80,garishpan90}.
Equivalently, the IRT is given by the coefficient in the linear term in the low-frequency expansion of
the susceptibility. 
For the uniaxial spins in a longitudinal field, the low-frequency perturbation theory
for the FPE leads to the analytical solution for the IRT in the whole temperature
range.
For more complicated models, such as a uniaxial spin in an oblique field, the
low-frequency susceptibility can be obtained numerically from the
matrix-continued-fraction solution of the FPE \cite{cofcrokalwal95}.
This routine is much more convenient than diagonalization of large matrices to obtain 
$\Lambda_1$.

Numerical solutions for the $\tau_{\rm int}$ and $\Lambda_1$ performed in 
ref.\ \cite{cofcrokalwal95} for the uniaxial spin model in a longitudinal field
revealed a striking feature.
Whereas in zero field $\tau_{\rm int}^{-1}$ coincides with $\Lambda_1$ in the whole
temperature range to within only 1.2\%, both quantities drastically deviate from each
other at low temperatures for the fields exceeding some critical value.
Whereas $\Lambda_1$ is exponentially small, as it should be, $\tau_{\rm int}^{-1}$
in not exponentially small and {\em increases} with the increase of the strength of 
the potential relief.
The physical explanation of this effect was given in ref.\ \cite{gar96pre}. 
In the biased low-temperature case, almost all particles or spins concentrate in the
lower well, whereas the upper well is depleted.
Then $\tau_{\rm int}$ captures the fast relaxation in the lower well (the intrawell
process) rather than the slow overbarrier relaxation process.
Since the intrawell process is not what one is usually interested in, the results of
refs.\ \cite{cofcrokalwal95,gar96pre} put the usefulness of
the concept of the integral relaxation time in a general case under question.

The aim of this Letter is to introduce another type of the IRT which describes the
process of the overbarrier relaxation in all situations.
The key idea is to consider the responce of the population difference 
$\Delta\equiv N_+-N_-$, where $N_\pm$ are the numbers of particles in the two wells,
 to the changing bias field. 
Unlike the magnetization response used to calculate the IRT, the $\Delta$ response is
insensitive to the redistribution of spins within the wells and always describes
the overbarrier process.
To emphasize the difference, it is convenient to designate the old (magnetization) IRT
as IRT-$M$ and the new (population) IRT as IRT-$N$.

Let us consider a single-domain magnetic particle with a classical Hamiltonian
%
\begin{equation}\label{ham}
\beta {\cal H} = v({\bf n}) - {\bf n}\cdot \delta \mbox{\boldmath$\xi$}(t,\omega),
\qquad \delta \mbox{\boldmath$\xi$}(t,\omega) = 
\delta \mbox{\boldmath$\xi$}_0 \exp(-i\omega t),
\end{equation}
where $\beta\equiv 1/T$, ${\bf n}$ is the unit vector representing the direction of
the particle's magnetization, 
$\delta \mbox{\boldmath$\xi$}(t,\omega) \equiv \beta VM_s \delta H(t,\omega)$ is the
reduced alternating field acting on the particle, $V$ is the particle's volume and
$M_s$ is the saturation magnetization.
The all-temperature analytical solution for the integral relaxation time is possible
if $v({\bf n})$ is axially symmetric and the alternating field is applied along the
$z$ axis.
The mostly studied model is a uniaxial particle in the longitudinal field
%
%
\begin{equation}\label{vxial}
v(x) = -\xi x - \sigma x^2, \qquad x\equiv \cos\theta 
\end{equation}
with
%
%
\begin{equation}\label{xiald}
\xi \equiv \beta VM_sH, \qquad \sigma \equiv \beta VM_s^2K,
\end{equation}
where $K>0$ is the anisotropy constant creating a potential relief with two wells
separated by a barrier.
The top of the barrier corresponds to
%
%
\begin{equation}\label{xm}
x = x_m = -h, \qquad h \equiv \xi/(2\sigma) = H/(2KM_s) ,
\end{equation}
the barrier height $\Delta v \equiv v(x_m) - v(-1)$ is given by 
%
%
\begin{equation}\label{delu}
\Delta v  = \sigma - \xi + \xi^2/(4\sigma) = \sigma (1-h)^2 ,
\end{equation}
for $h\geq 1$ the barrier disappears.

The dynamics of the classical single-domain magnetic particle is governed by the
Fokker-Planck equation which was first formulated by Brown \cite{bro6379}.
This FPE can also be found in refs.\ \cite{garishpan90,gar96pre,garkencrocof99prl}.
The equilibrium distribution function satisfying the static FPE can for axially
symmetric models be written in the form
%
%
\begin{equation}\label{f0z}
f_0(x) = \frac{\exp[-v(x)]}{Z}, \qquad Z = \int_{-1}^1 \exp[-v(x)] \drm x .
\end{equation}
If an infinitesimally small alternating longitudinal field is applied,
the distribution function can be searched in the form
%
%
\begin{equation}\label{df}
f(x,t,\omega) = f_0(x)\big[1 + q(x) \delta \xi(t,\omega)\big],
\end{equation}
where the complex function $q(x)$ satisfies the equation \cite{garishpan90,gar96pre}
%
%
\begin{equation}\label{qeq}
\left( \frac{\drm}{\drm x} - \frac {\drm v}{\drm x} \ \right)
(1-x^2)\frac{\drm q}{\drm x} + 2i\omega \tau_N q 
= -(1-x^2)\frac {\drm v}{\drm x} - 2x .
\end{equation}
Here $\tau_N$ is the N\'eel relaxation time defined by
%
%
\begin{equation}\label{lamn}
\tau_N^{-1} \equiv \Lambda_N  \equiv 2\gamma a T/(VM_s) ,
\end{equation}
where $\gamma$ is the gyromagnetic ratio and $a$ is the dimensionless Gilbert's
damping parameter.
The solution of eq.\ (\ref{qeq}) should satisfy the normalization condition
$\int_{-1}^1 f_0(x) q(x) \drm x =0$.
The low-frequency expansion of $q(x)$ has the form
%
\begin{eqnarray}\label{qxsol}
&&
q(x) = q_0(x) + i\omega\tau_N q_1(x) + \ldots,
\qquad q_0(x) = x-B
\nonumber \\
&&
q_1(x) = \int_{-1}^x p(x') \drm x' - \int_{-1}^1 f_0(x) \int_{-1}^x p(x') \drm x' \drm x,
\end{eqnarray}
where $B = \int_{-1}^1 x f_0(x) \drm x$ is the equilibrium spin polarization and
%
%
\begin{equation}\label{pxsol}
p(x) = \frac 2 { 1-x^2 } \frac {\Phi(x)}{f_0(x)},
\qquad \Phi(x) = \int_{-1}^x (B-x') f_0(x') \drm x'.
\end{equation}

Having solved the FPE, one can calculate the dynamic susceptibility
%
%
\begin{equation}\label{chiM}
\tilde\chi(\omega) = \frac 1{M_s} \frac{ \delta M(t,\omega)}{\delta \xi(t,\omega)}
= \int_{-1}^1 x f_0(x) q(x) \drm x 
\cong \tilde\chi (1+i\omega\tau_{\rm int} + \ldots), 
\qquad \tilde\chi = B' \equiv \frac {\partial B}{\partial \xi}.
\end{equation}
%
%
Using eq.\ (\ref{qxsol}) and integrating by parts one obtains the magnetization IRT
\cite{sza80,garishpan90,gar96pre}
%
%
\begin{equation}\label{tauint}
\tau_{\rm int} = \frac{2\tau_N}{B'} 
\int_{-1}^1 \frac{\drm x}{1-x^2} \frac{\Phi^2(x)}{f_0(x)} .
\end{equation}
In ref.\ \cite{gar96pre} it was shown that in the biased case $\xi >0$ at low
temperatures the integrand of this formula peaks near the top of the barrier $x=x_m$
and in the lower well $x \approx 1$, which corresponds to the overbarrier and intrawell
processes, respectively.
For $h > h_c = 3-2\sqrt{2} \approx 0.17$ at low temperatures, the upper well is
depleted, the overbarrier process is thus suppressed, and the intrawell process controls the IRT.    
This undesirable feature makes one to look for a more appropriate definition of the
integral relaxation time.

\begin{figure}[t]
\unitlength1cm
\begin{picture}(7,8)
\centerline{\epsfig{file=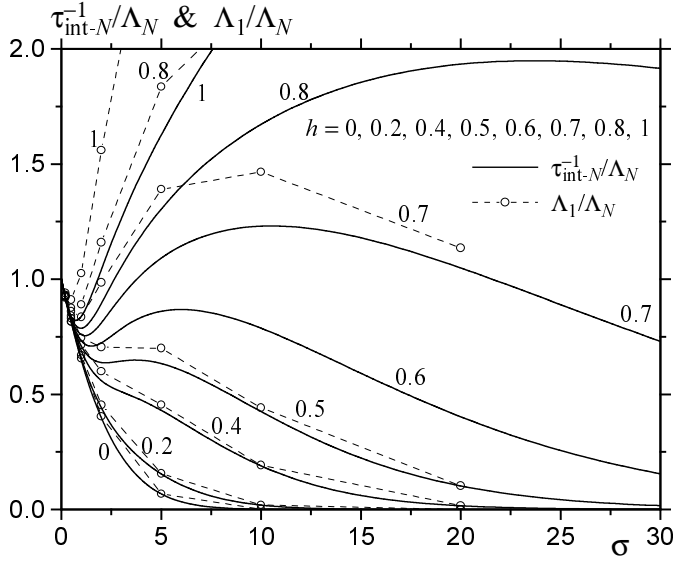,angle=-90,width=10cm}}
\end{picture}
\caption{ \label{taun}
Inverse of the integral relaxation time for populations IRT-$N$, 
eq.\ (\protect\ref{tauintN}), and the lowest
eigenvalue of the Fokker-Planck equation $\Lambda_1$ \protect\cite{cofcrokalwal95} vs the potential strength $\sigma$.
}
\end{figure}

The simplest choice for the quantity which is insensitive to the intrawell process is
the difference of the wells populations
%
%
\begin{equation}\label{DeltaDef}
\Delta(t,\omega) \equiv N_+ - N_- = \int_{-1}^1 {\rm sgn}(x-x_m) f(x) \drm x.
\end{equation}
In equilibrium one has
%
%
\begin{equation}\label{DeltaEqu}
\Delta = \frac{ Z_+ - Z_-}{Z_+ + Z_-}, \qquad
Z_- = \int_{-1}^{x_m}  f_0(x) \drm x,
\qquad Z_+ = \int_{x_m}^1  f_0(x) \drm x,
\end{equation}
where $Z_\pm$ are the contributions to the partition function $Z$ from both wells.
The dynamic susceptibility for the population difference $\Delta$ has the form
%
%
\begin{equation}\label{chiN}
\tilde\chi_\Delta(\omega) = \frac{ \delta \Delta(t,\omega)}{\delta \xi(t,\omega)}
= \int_{-1}^1 {\rm sgn}(x-x_m) f_0(x) q(x) \drm x 
\cong \tilde\chi_\Delta (1+i\omega\tau_{{\rm int},N} + \ldots), 
\end{equation}
where    
%
%
\begin{equation}\label{DerDel}
\tilde\chi_\Delta = \Delta' \equiv \frac {\partial \Delta}{\partial \xi }
= \frac{ 2(Z_- Z_+' - Z_+ Z_-')}{(Z_+ + Z_-)^2},
\qquad Z_\pm' \equiv \frac {\partial Z_\pm}{\partial \xi },
\end{equation}
and the integral relaxation time for the populations is given by
%
%
\begin{equation}\label{tauintN}
\tau_{{\rm int},N} = \frac{2\tau_N}{\Delta'} 
\int_{-1}^1 \frac{\drm x}{1-x^2} \frac{\Phi_N(x)\Phi(x)}{f_0(x)} ,
\qquad \Phi_N(x) = \int_{-1}^x [\Delta -{\rm sgn}(x'-x_m)] f_0(x') \drm x'.
\end{equation}
This result follows from the same low-frequency solution of the FPE, eq.\
(\ref{qxsol}).
It can be checked that the function $\Phi_N(x)$  has no maxima near the bottom
of the wells, thus $\tau_{{\rm int},N}$ does not describe the intrawell process. 
This new integral relaxation time, which can be abbreviated as IRT-$N$, captures the
overbarrier process in the low-temperature range for all values of the longitudinal
field.

Dependences of $\tau_{{\rm int},N}^{-1}$ on the barrier-strength parameter $\sigma$ for
different values of $h$ for the uniaxial model 
[see eqs.\ (\ref{vxial}) and (\ref{xiald})] are shown in Fig.\ \ref{taun},
compared to the values of the lowest eigenvalue $\Lambda_1$ quoted in Table I of 
ref.\ \cite{cofcrokalwal95}.
One can see that both quantities behave similarly, in contrast to 
$\tau_{\rm int}^{-1}$ (cf. Figs. 1 and 2 of ref.\ \cite{cofcrokalwal95}). 
In the region where the escape rate becomes exponentially small, 
$\tau_{{\rm int},N}^{-1}$ and $\Lambda_1$ coincide with each other.
An interesting feature of the dependences in Fig.\ \ref{taun} is the maximum of the
relaxation rate arising in the strongly biased case, which is more pronounced for 
$\Lambda_1$ than for $\tau_{{\rm int},N}^{-1}$.
The physical reason for this maximum is the following.
As long as the parameters $\sigma$ and $\xi$ are large but the reduced barrier height
$\Delta v = \sigma(1-h)^2$ is small, the dynamics of the magnetic particle is qualitatively the
same as that of a magnetic particle {\em without} a potential barrier at low
temperatures.
In this case the relaxation rate can be found in a simple way from the deterministic
Landau-Lifshitz equation and it is proportional to the strength of the potential
relief, i.e., to $\sigma$.
If $\sigma$ is further increased, then for any $h<1$ the reduced barrier hight 
$\Delta v$ becomes large for $\sigma \gg 1/(1-h)^2$, and the escape rate becomes
exponentially small.   
These two regimes are separated by a maximum of the relaxation rate.

In this Letter, the integral relaxation time for populations, IRT-$N$, was introduced.
This quantity is more useful than the previously considered integral relaxation time
for the magnetization, IRT-$M$, and it always captures the exponentially slow process
of thermal activation in the high-barrier limit.
For the simplest models considered here, there is an all-temperature solution for the 
IRT-$N$ in quadratures. 
In other cases, the problem can be solved numerically considering the low-frequency
limit of the matrix-continued-fraction solution of the FPE \cite{cofcrokalwal95}.
Formulation of the method for non-axially symmetric potentials or for (mechanical)
particles poses no problems.
Numerical calculation of the IRT-$N$ is a more convenient and accurate procedure than
that for the lowest eigenvalue $\Lambda_1$. 
Using the new integral relaxation time should help to throw a light on fine details of
various crossovers which have been recently discovered for the uniaxial model in a
transverse field \cite{garkencrocof99prl}.
For this model, numerical calculation of $\Lambda_1$ faces difficulties in the
low-damping high-barrier case.

The author thanks Oleg Yevtushenko for critical reading of the manuscript.


\end{document}